\documentclass[conference,dvipsnames]{IEEEtran}

\usepackage{bm}
\usepackage{amsmath,amssymb,amsfonts}
\usepackage{pifont}
\usepackage{algcompatible}
\usepackage[ruled]{algorithm2e}
\usepackage{algpseudocode}
\usepackage{steinmetz}
\usepackage{graphicx}
\graphicspath{{sims/}}
\usepackage{textcomp}
\usepackage{color,xcolor}
\usepackage{lipsum}
\usepackage{array}
\usepackage{multirow}
\usepackage[colorlinks=true, linkcolor=OliveGreen, urlcolor=Fuchsia, citecolor=blue, breaklinks=true]{hyperref}
\usepackage{xurl}
\usepackage{adjustbox}
\usepackage{fancyvrb}
\usepackage{booktabs}
\usepackage{balance}
\usepackage[colorlinks=true, linkcolor=OliveGreen, urlcolor=Fuchsia, citecolor=blue,breaklinks=true]{hyperref}

\begin{document}

\title{Simulation of Custom-Precision OCP MX Block Floating-Point Formats and Arithmetic}

\author{\IEEEauthorblockN{Maliha Islam and Mantas Mikaitis}
  \IEEEauthorblockA{School of Computer Science, University of Leeds, Leeds, UK
  }}

\maketitle

%
%
\begin{abstract}
  We present MXsim v0.1: a MATLAB library for simulating the OCP MX block floating-point arithmetic and formats.
  The library is built on top of a custom-precision floating-point simulator, CPFloat, which allows custom variants of non-block IEEE 754 floating-point formats to be simulated.
  MXsim is a MATLAB simulator, and is targeted at mixed-precision algorithm development community to perform experimentation in OCP MX block floating-point arithmetic by using a simple and well-documented interface in MATLAB, which is especially useful when access to the latest NVIDIA and AMD GPUs is not available.
  Furthermore, due to full customisation of exponents, precisions of input and output formats, and the precision of the accumulator within the dot product operation, MXsim allows experimenting with custom variants of the OCP MX block floating point for research.
  Finally, since the bit-wise numerical behavior of OCP MX formats is not documented by GPU vendors, and the precision and rounding are not prescribed by the OCP specification itself, MXsim allows experiments with various possible settings in order to determine the effects on mixed-precision numerical algorithms.
\end{abstract}

\section{Introduction}

Computing hardware is increasingly incorporating low- and mixed-precision floating-point formats and arithmetic, motivated by the need to increase efficiency, whilst reducing memory usage and hardware complexity. In 2023, OCP released two floating-point format specifications, in which they standardised 4-8 bit representations, with a framework similar to block floating-point (BFP), named \emph{microscaling (MX) formats}~\cite{ocp23a, ocp23b}.
These formats are now present in NVIDIA Blackwell~\cite{nvid25} and AMD CDNA4-based GPUs \cite{amd25b}.
Because access to these GPUs is scarce we have created a simulator MXsim\footnote{\url{https://github.com/north-numerical-computing/MXsim}} for the OCP MX formats which will be useful to those wanting to experiment with OCP MX easily and to those who want to experiment with MX-like block floating-point arithmetic in general.
Our own motivation is, in subsequent work, to use the library for research on algorithms for emulating high-precision computations with the MX formats.

BFP is an alternative to standard floating point which does not appear commonly in literature but has recently been appearing in hardware accelerators targeted at machine learning. It is now present in commercial hardware and has been standardised by OCP, as well as IEEE in the form of the upcoming P3109 standard \cite{P3109_Public}. BFP hardware could allow for the use of more-efficient low-precision arithmetic, whilst preserving accuracy, in scientific computing.

Although our library is inspired by the OCP MX specification \cite{ocp23b} that contains five block floating-point formats, the exponents, precisions of input and output formats, and the precision of the accumulator within the dot product operator can be customised, allowing for the simulation of many more than five custom MX-like BFP formats.

\section{Previous work}

MXsim has CPFloat \cite{cpfloat} as a dependency, a MATLAB library for simulating custom floating-point formats. CPFloat provides a general-purpose IEEE 754 floating-point format simulation that supports customisable precisions, exponent widths, and rounding modes. Our software in effect extends CPFloat, providing MX block floating-point structures, with shared scale factors, as well as operations on these block formats.

A software called gfloat~\cite{gfloat} provides generic floating-point encode and decode logic in Python, supporting OCP MX formats alongside IEEE 754 and P3109 formats.
However, like CPFloat, it does not implement MX
operations such as the dot product or matrix multiplication of two vectors
of floating-point values represented in user-specified formats, scaled by block scales.

Microsoft's MX PyTorch emulation library \cite{microxcaling} provides emulation of MX-compatible formats in Python. It requires a CUDA-capable GPU and a PyTorch installation, and supports a fixed set of OCP MX formats (see. Section~\ref{sec:ocpmx}) without the ability to customise element precision, block size, or accumulation format.
Hence, this emulation library is targeted at machine learning users and supports only currently-available OCP formats.
In contrast, MXsim targets the scientific computing community, providing full customisation beyond the standard OCP formats and can run on any hardware that supports MATLAB.

Several hardware implementations of MX formats have been proposed.
Samson et al. \cite{samson24} present an open-source FPGA
implementation of the full MX standard, supporting all concrete formats for conversion and the standard-defined operations.
Gorodecky and Sousa \cite{gorodecky24} propose an FPGA-based hardware converter from IEEE 754 fp32 to MX formats, supporting six MX format variants---it includes one MX format that uses fixed-point arithmetic for elements of vectors.
They implement the conversion algorithm specified in the OCP MX specification \cite[Sec.~6.3]{ocp23b}.  Regarding exceptional values, they note that the shared scale $X$ (see Section~\ref{sec:ocpmx}) can represent NaN but not infinity.
If $X$ is a NaN, the vector represented by the MX vector with $X$ a NaN contains elements that are all NaNs.
This is not specified by OCP, but is derived from Equation~\ref{eq:blockvect}.

Abdurakhmanov and Fahmy \cite{abdurakhmanov25} present a detailed exploration of systolic array architectures for MX format matrix multiplication on FPGAs, evaluating two accumulation strategies across multiple MX formats and pipeline depths. One of these accumulation strategies is to generate exact results, and the following formula is given to determine the bitwidth required for exact accumulation:

\begin{equation}
\label{eq:accumulation}
    \text{accwidth} = 2 \times (2^{E_{\text{width}}} + M_{\text{width}}) 
    + \lceil \log_2(k) \rceil
\end{equation}
where $E_{\text{width}}$ is the number of exponent bits and $M_{\text{width}}$ is the number of stored mantissa bits (i.e. the fractional part of the significand $m$ defined in Section~\ref{sec:ieee754}, excluding the implicit leading bit), and $k$ is the block size.

For the OCP 8-bit formats (see. Section~\ref{sec:ocp8bit}), the minimum bit widths of the accumulator are 73 and 43 for element formats E5M2 and E4M3, respectively. This formula is also used by Samson et al. \cite{samson24}.

The second accumulation strategy proposed by \cite{abdurakhmanov25} is BFloat16 accumulation, which provides better performance than exact accumulation. BFloat16 is chosen over fp16 due to its exponent being 8 bits wide, allowing for a larger dynamic range than E5M2.

Islamoglu~et~al.~\cite{islamoglu25} and Wipfli~et~al.~\cite{wipfli26} propose RISC-V ISA extensions for native MX dot product execution, achieving significant speedup over software emulation.
This work is targeted at hardware implementations and does not contribute to the research in applied mathematics, where simulation of new formats in an easy-to-use manner is a very useful tool for algorithmic research and numerical experiments.
MXsim provides a software tool for experimentation with MX arithmetic without requiring access to specialised hardware.

\section{Background}

\subsection{IEEE 754 floating-point formats}
\label{sec:ieee754}
A floating-point number in a binary format with precision $p$ and 
exponent range $[e_{\min}, e_{\max}]$ is represented as:
\begin{equation}
    x = (-1)^s \cdot m \cdot 2^e
\end{equation}
where $s \in \{0, 1\}$ is the sign bit, $m$ is the significand 
satisfying $1 \leq m < 2$ for normal numbers ($0 \leq m < 1$ 
for subnormal numbers), and $e \in [e_{\min}, e_{\max}]$ is the 
exponent. The exponent range satisfies $e_{\min}=1-e_{\max}$~\cite{ieee19}. The significand $m$ has $p$ bits, giving $p-1$ stored 
mantissa bits. The IEEE 754 standard~\cite{ieee19} defines 
several binary formats, including fp32 (binary32, $p = 24$, $e_{\min} = -126$, $e_{\max} = 127$) and 
fp64 (binary64, $p = 53$, $e_{\min} = -1022$, $e_{\max} = 1023$).

\subsection{Definition of OCP 8-bit formats}
\label{sec:ocp8bit}
The OCP 8-bit floating-point specification \cite{ocp23a} defines 
two formats: E4M3 and E5M2. Both follow the general floating-point representation of Section~\ref{sec:ieee754}, with a sign bit, exponent bits, and significand bits, including an implicit bit and subnormals. 

E4M3 represents the exponent $e$ in 4 bits and the fractional part of the significand $m$ in 3 bits, with $p = 4$, $e_{\min} = -6$ and $e_{\max} = 8$, giving a maximum finite value of 448. 

E5M2 represents the exponent $e$ in 5 bits and the fractional part of the significand $m$ in 2 bits, giving a total precision $p = 3$, with $e_{\min} = -14$ and $e_{\max} = 15$, providing a larger dynamic range at the cost of reduced precision. Note that E4M3 violates $e_{\min} = 1 - e_{\max}$.

\subsection{Definition of OCP MX formats}
\label{sec:ocpmx}
An MX-compliant format consists of a block of $k$ private elements, where each element $P_i$ is a floating-point number, and a shared scale factor $X$. All elements $P_i$ must have the same floating-point format and $X$ must be a power of 2. Each value $V_i$ within the block is given by
\begin{equation}
\label{eq:blockvect}
    V_i = X \cdot P_i.
\end{equation}
When converting to MX, the shared scale $X$ is computed by taking the largest power of two less than or equal to the largest absolute value within the block, divided by the largest power of two representable in the element format \cite{ocp23b}. 

The specification defines five floating-point variants (MXFP8 E5M2/E4M3, MXFP6 E3M2/E2M3, MXFP4 E2M1) \cite{ocp23b} and one integer variant (MXINT8),
all with a fixed block size of $k = 32$. MXsim focuses on floating point element formats, but may be extended in the future to include MXINT8 if there is demand for it.

\subsection{Differences of OCP formats compared with IEEE 754 formats}
In IEEE 754, each number carries an exponent part of its bit pattern. In MX formats, each element similarly retains its own exponent, but additionally shares a common scale factor $X$ across a block of $k$ elements.

In IEEE 754 style formats, the maximum exponent bit pattern is reserved for representing infinities and multiple NaN values. In E4M3, this bit pattern is instead used for finite numbers, with only one significand pattern (all-ones) reserved for NaN. This change causes a violation of the IEEE property $e_{\min} = 1 - e_{\max}$, and gives one more finite exponent value than an equivalent IEEE style format would have. The maximum finite value is $2^8 \times (1 + 6/8) = 448$ \cite{ocp23a}. If E4M3 had followed the IEEE style for that last binade, the maximum value would be $2^8 \times (1 + 7/8) = 480$ since the all-ones significand would not be reserved for the NaN.

Similar to IEEE 754~\cite[Sec.~9.4]{ieee19}, the OCP specification leaves the internal precision and order of operations of the dot product implementation-defined~\cite{ocp23b}.

\section{Methods}
\label{sec:methods}
\subsection{Functional requirements}
    \subsubsection{\textbf{Format selection}}
    Formats can be selected through a structure defined by the user. We will refer to this structure as \texttt{MX}.
    The \texttt{MX} struct configures the block floating-point format. 
    The element type is set via \texttt{element\_opts}, with block size
    \texttt{k}. Accumulation and output formats are configured by structs
    \texttt{accumulator\_opts} and \texttt{output\_opts} respectively,
    and the shared scale is parameterised by \texttt{scale\_bits}. These options allow for full precision customisation of the MX structure and its operations.
        If a field of \texttt{MX} is left undefined, it will default to its corresponding value denoting OCP MXE4M3~\cite{ocp23b}.

    \begin{table}[h]
      \centering
      \caption{Fields of the \texttt{MX} configuration struct.}
      \begin{tabular}{llll}
        \toprule
        \textbf{Field} & \textbf{Type} & \textbf{Default}  \\
        \midrule
        \texttt{MX.element\_opts} & struct & E4M3 \\
        \texttt{MX.k}             & integer & 32  \\
        \texttt{MX.accumulator\_opts} & struct & fp32 \\
        \texttt{MX.output\_opts}  & struct & fp32 \\
        \texttt{MX.scale\_bits}   & integer & 8   \\
        \bottomrule
      \end{tabular}
      \vspace{0.5em}
      \label{tab:mx_fields}
    \end{table}

    As shown in Table~\ref{tab:mx_fields}, some of the fields of 
    MX are structures, each of which can specify any floating-point format supported by CPFloat~\cite{cpfloat}. CPFloat provides a
    number of predefined formats (including E4M3 and E5M2), as well as support for custom formats---this is exploited in MXsim to provide customisation in the simulation of MX formats.

    \subsubsection{\textbf{Accumulation}}
    The accumulation precision for MX operations is unspecified by OCP. 
    The default accumulation format is fp32, which is provided by CPFloat \cite{cpfloat}. CPFloat uses fp64 as its base arithmetic and rounds to the specified precision, which limits the maximum precision it can support to 26 bits due to double rounding effects \cite{rump2023precision}. Consequently, exact accumulation for the OCP 8-bit formats, which requires 43 and 73 bits for E4M3 and E5M2 respectively (see Equation~\ref{eq:accumulation}), cannot be reliably achieved through CPFloat. Even native fp64 accumulation, which would in principle provide exact results for E4M3, cannot be enabled via CPFloat.
    
    Future work will explore providing an option for fixed-point accumulation, by using an alternative to CPFloat.

    Alternatively, for non-exact accumulation, Abdurakhmanov~and~Fahmy~\cite{abdurakhmanov25} favour the use of BFloat16, due to its 8-bit exponent, offering a wider dynamic range than fp16 (IEEE binary16 format).

    Recent work characterising the numerical behaviour of NVIDIA B200  Tensor Cores \cite{khattak2025tensorcore} found that the accumulator for 8-bit formats uses 25-bit fixed-point arithmetic with 2 integer bits. It is highly likely that MX hardware will share a similar accumulation strategy. In future work, we plan to set the default accumulator to match this precision, using a dedicated fixed-point model, as CPFloat does not currently support fixed-point arithmetic.

    The accumulation format in MXsim can be configured by the user, via \texttt{MX.accumulator\_opts}.

    \subsubsection{\textbf{Conversion}}
    Once \texttt{MX} has been defined, a vector $V$ of any length can be converted to the chosen format via the function \texttt{MXconvert(V,MX)}. The function will return a structure, with the field \texttt{scale} containing a set of shared scales and the field \texttt{vector} containing a set of vectors of length \texttt{k}.
    If the input vector is not divisible by $k$, it will first be padded with zeros, so that its length is a multiple of k, before being split into blocks. 
   
    The scale and vector components of each block are calculated by the internal block convert function \texttt{BlockConvert}, which is an implementation of Section 6.3 of the OCP MX specification \cite{ocp23b}. The scale $X$ is computed for each block as described in Section~\ref{sec:ocpmx}. The input vector is then divided by $X$ and converted to the format specified by \texttt{element\_opts}, using CPFloat, to give the private elements $P_i$.
    If the element format is E4M3, the maximum value of 448 is forcibly set within the conversion algorithm, as CPFloat has a maximum value of 480.

    The conversion to the element format uses round-to-nearest with ties to even by default (see. Section~\ref{sec:round}), but this is configurable via \texttt{element\_opts.round}. On overflow, values exceeding the maximum finite value of the element format are clamped to the maximum by default, as specified by OCP \cite{ocp23b}. This behaviour can be configured via \texttt{MX.element\_opts.saturation}. Further detail on this can be found in Section~\ref{sec:exceptions}.

    MXsim also contains the function \texttt{MXconvert\_inv(M,n)},
    which converts a vector from MX format to standard floating point.
    This function is designed to be the inverse of \texttt{MXconvert}.
    The argument \texttt{M} must be an MX structure, containing a field of vectors and a field of scales.
    The argument $n$ is optional, and is the length of the original vector. This argument reverses the effects of the zero padding which occurs during conversion, if the original vector length is not a multiple of $k$, and sets the length of the backwards converted vector to $n$.
    The function multiplies each block vector by its corresponding scale and concatenates the results into a single resulting vector. If $n$ is not supplied, the resulting vector will be a multiple of $k$, which may not match the length before conversion to MX. 

    \subsubsection{\textbf{Operation support}}
    There are two operations specified by OCP \cite{ocp23b}. 

    The dot product of two single blocks of length k is defined \cite[Sec.~6.1]{ocp23b} as
    \begin{equation}
    \mathrm{Dot}(A, B) = X^{(A)} X^{(B)} \sum_{i=1}^{k} P_i^{(A)} P_i^{(B)},
    \label{eq:dot}
  \end{equation}
  where $X^{(A)}$ and $X^{(B)}$ are the shared scales of blocks $A$ and $B$ respectively (see. Section~\ref{sec:ocpmx}), and $P_i^{(A)}$, $P_i^{(B)}$ are the $i$-th floating-point elements of each block.

  Operation~\ref{eq:dot} is supported by the function \texttt{BlockDot(A, B, MX)}, and computes the dot product of a single block. The input values \texttt{A} and \texttt{B} should be in MX format, and with vector components of length \texttt{k}. The accumulation precision is determined by \texttt{MX.accumulator\_opts}. The sum is currently computed recursively, which may not reflect the accumulation behaviour of NVIDIA hardware. In future work we plan to utilise the accurate models of NVIDIA Tensor Cores presented in \cite{khattak2025tensorcore} to provide an option for hardware-accurate accumulation.

     The second operation specified in OCP MX Section 6.2  \cite{ocp23b} is the dot product of a vector of any length:
     \begin{equation}
     C = \mathrm{DotGeneral}(A, B) = \sum_{j=1}^{n} \mathrm{Dot}(A_j, B_j)
     \label{eq:dotgeneral}
    \end{equation}
    where $A_j$ and $B_j$ are the $j$-th blocks of vectors $A$ and $B$ respectively, each of length $k$.

    Operation~\ref{eq:dotgeneral} is supported by the function \texttt{GeneralDot(A,B,MX)}. Inputs \texttt{A} and \texttt{B} are MATLAB vectors of any length. The sum is computed recursively, using the precision defined by \texttt{MX.output\_opts}.
 
    MXsim also contains a matrix multiplication operation, \texttt{BlockMatMul(A,B,MX)} which is not present in the OCP specification. The arguments \texttt{A} and \texttt{B} are multiplicable matrices of sizes $m \times n$ and $n \times p$ respectively. The result is computed via a sum of inner products computed by \texttt{GeneralDot}, iterating over rows of $A$ and columns of $B$, accumulated recursively.

    \subsubsection{\textbf{Rounding mode}}
    \label{sec:round}
    Round to nearest with ties to even is used by default at each of the three sources of rounding: when rounding inputs to the element format, when rounding during accumulation, and when rounding the accumulator to the output format. Each of these can be customised via the corresponding options in the \texttt{MX} struct, using the CPFloat rounding options accessible via \texttt{help cpfloat}.

\subsection{Treatment of exceptional values}
\label{sec:exceptions}

     \subsubsection{\textbf{Exponent limit}}
     The shared scale factor $X$ in OCP-defined MX formats is E8M0 and contains only exponent bits. E8M0 has one representation for NaN: the all-ones bit pattern. It does not have a representation for infinity, as all other bit patterns are used for finite values.  
     E8M0 is set as the default format for $X$ in this software, but this is configurable via \texttt{MX.scale\_bits}. This field must be set to an integer, as it determines the bitwidth of the scale. The scale will contain an exponent part, with zero significand bits, as it must always be a power of two \cite{ocp23b}.

     All scale formats will mirror E8M0, with a single NaN representation and no infinity representation. This ensures that the exception handling within blocks stays consistent, regardless of shared scale bitwidth. 
     
     If the computed exponent exceeds the maximum representable scale exponent, $X$ is clamped to its maximum value $2^{E_{\max}}$, where $E_{\max} = 2^{s-1} - 1$ and $s$ is the number of scale bits.

    \subsubsection{\textbf{Infinities and saturation}}
    
    As mentioned, the exponent part of a block can not be set to infinity. If a vector containing infinities is converted to MX, the shared exponent within its block will be set to its maximum value.
    Treatment of infinities within vector parts of a block can be configured by the user, using the \texttt{saturation} field of \texttt{MX.element\_opts}. By default, $\texttt{saturation} = 1$,
    clamping all values exceeding the element maximum, including infinities, to the element maximum. This default was chosen as it is specified in the OCP MX conversion algorithm \cite{ocp23b}.
    If the user sets \texttt{saturation} to $0$, infinities may be present in the block vector, and will remain in their corresponding position before conversion. If the element format is E4M3, infinities in the vector will be represented as NaN, in accordance with OCP \cite{ocp23a}. 

    If a vector contains only NaNs and infinities, the scale is computed from the infinite values, resulting in the maximum representable scale. NaN positions are preserved in the element vector. This ensures that Inf and NaN places from the original vector are preserved when using \texttt{MXconvert\_inv}, whilst also ensuring a NaN outcome if \texttt{BlockDot} is used.

    \subsubsection{\textbf{NaNs}}
    \label{sec:nans}
    The OCP specification mentions that if the shared scale is NaN, then a backwards conversion would result in an all-NaN vector \cite{ocp23b}, and MXsim preserves this behaviour.
    However, the handling of a vector containing NaNs when converting to MX is left as implementation defined by OCP.

    When a vector containing a NaN is converted to MX via \texttt{MXconvert}, the shared scale $X$ will be calculated based on the non-NaN values within the vector. The NaNs in the resulting block will preserve their corresponding positions in the vector part of the MX format. This is consistent with Samson et al. \cite{samson24}, which mentions setting NaNs at the element level if possible.
    
    If the input vector contains only NaNs, it is not possible to compute a finite shared scale, as there are no finite values to compute it from. Hence, the scale is set to NaN, as well as all the elements. 

    If a dot product is performed on any MX-vector containing a NaN, the resulting value will be a NaN.

    An alternative approach would be to set the shared scale to NaN, if any NaNs are present, resulting in a vector part of all NaNs. We chose not to use this method, as it does not preserve the other values within the vector when backwards converting, and sets them all to NaN. The chosen method preserves NaN positions as well as finite values (up to conversion error), while still producing NaNs after a dot product. 

 \subsubsection{\textbf{Zero vectors}}
 If all elements of a block are zero, the shared scale is set to 1 and the private elements are set to zero. This avoids the undefined behaviour that would result from computing $\lfloor \log_2(0) \rfloor$ in the scale computation. This behaviour is left implementation defined by OCP \cite{ocp23b}. 
    
    \subsubsection{\textbf{Subnormals and negative zero}}
    The OCP MX specification requires support for subnormal numbers in FP8 formats \cite{ocp23a}, with minimum subnormal magnitudes of $2^{-9}$ and $2^{-16}$ for E4M3 and E5M2 respectively. These are handled by CPFloat \cite{cpfloat}. 
    
    Negative zero is preserved through conversion, which is consistent with the OCP specification \cite{ocp23a}.

\subsection{Structure of the MATLAB simulator}

  The repository is organized into \texttt{MXsim/} and \texttt{tests/} directories, with CPFloat \cite{cpfloat} as an external dependency that must be installed separately and added to the MATLAB search path.

  \subsubsection{\textbf{\texttt{simulator/}}}
  The core simulator functions are located in \texttt{simulator/}. The function \texttt{MXconvert} converts a vector of any length to MX format, returning a structure containing block vectors and their corresponding scales. The internal function \texttt{BlockConvert} implements the OCP conversion algorithm for a single block \cite{ocp23b}. Operations are provided by \texttt{BlockDot}, \texttt{GeneralDot}, and \texttt{BlockMatMul}, which compute the block dot product, general dot product, and matrix multiplication, respectively. Inverse conversion back to standard floating-point is provided by \texttt{backwardsConvert}, which operates on a single block, and \texttt{MXconvert\_inv}, which operates on a vector of any length. Default formats are set by \texttt{default\_MX}. Fields of \texttt{MX} unspecified by the user are set to their default values via \texttt{merge\_MX}.

  \subsubsection{\textbf{tests/}}
  The \texttt{tests/} directory contains integration tests in \texttt{tests/integration/} and error bound tests in \texttt{tests/expected\_errors/}. All tests can be run via \texttt{runalltests.m}.

\subsection{Testing of the correctness}

Since the documentation of MX-compliant hardware is not precise, 
particularly regarding the ordering of additions and internal precision of the accumulator, there is no gold standard reference implementation against which MXsim can be directly validated. Instead, correctness is verified through two approaches.

  \subsubsection{Integration tests} 
  Integration tests are located in \texttt{tests/integration/}, and cover the core functions \texttt{BlockConvert}, \texttt{BlockDot}, \texttt{MXconvert}, and \texttt{backwardsConvert}. These test edge cases and hand computed results against MXsim results.

  \paragraph{BlockConvert}
  \texttt{BlockConvert} is the core function behind MX conversions, and explicitly follows the OCP Section 6.3 conversion algorithm \cite{ocp23b}. Hence, it has been thoroughly tested. Tests include verification against a variety of hand computed results. The input vectors for the hand computed tests include the following: 
  \begin{itemize}
    \item Multiples of 2 in E4M3, verifying correct scale and quantised values.
    \item Prime numbers in E4M3, verifying correct handling of non-power-of-two values.
    \item Mixed-sign values in E4M3, verifying correct sign handling.
    \item Decimal values in E5M2, verifying correct quantisation.
    \item Uniform in E5M2, where all elements are equal, verifying all elements produce the same output.
    \item Large dynamic range values in E5M2, verifying correct scale selection.
  \end{itemize}
 In addition, edge cases and exception handling (see. Section~\ref{sec:exceptions}) have been tested. Each of the following criteria have been given a dedicated test: 
 \begin{itemize}
    \item The scale is always an exact power of two.
    \item The output vector length matches the input length.
    \item Values exceeding the E4M3 maximum of 448 are clamped under saturation mode.
    \item A vector containing infinities produces a finite scale.
    \item NaN positions are preserved in the output vector, with the scale computed from non-NaN values.
    \item An all-NaN vector produces a NaN scale.
    \item A zero vector produces a scale of 1 and a zero output vector.
\end{itemize}

  \paragraph{BlockDot}
  Hand-computed test cases verify the dot product of two MX blocks using the formula in Section 6.1 of the OCP MX specification \cite{ocp23b}. Two test cases are included: one using E4M3 format and one using E5M2 format, each with known scales and element vectors. fp16 accumulation was used for handwritten test cases. Edge cases tested include NaN handling from both the element vector and the scale factor.

  \paragraph{backwardsConvert}
  Tests verify correct reconstruction of values from a single MX block for a range of scale factors, including positive and negative powers of two, and vectors containing positive, negative, mixed-sign, and zero values. Edge cases include zero scale, single element blocks, and NaN handling from both the scale and the element vector.

  \subsubsection{Expected error tests}
  Error bound tests are located in \texttt{tests/expected\_errors/}, and verify that the results of \texttt{BlockDot} and \texttt{GeneralDot} fall within the worst-case forward error bound for recursive summation \cite[Eq.~4.4]{higham2002}.

  The vector length, vector values and unit roundoff are subbed into the theoretical bound at each iteration of this test, to find the maximum accumulation error. This is then plotted against errors from the randomly generated test cases, to show that the results fall within the expected range.

 \paragraph{BlockDotErrorBoundTest}
\label{sec:BlockDoterr}
The error bound for \texttt{BlockDot} is derived from the standard model of floating-point arithmetic \cite{higham2002}, in which each operation satisfies 
$fl(a \,\mathrm{op}\, b) = (a \,\mathrm{op}\, b)(1+\delta)$ with $|\delta| \leq u$, 
where $u = 2^{-24}$ is the unit roundoff of the fp32 accumulation format.
\texttt{BlockDot} computes $k$ products, and performs a recursive summation. Substituting these terms into Higham's equation for recursive summation \cite[eq.~4.4]{higham2002}, the error $E_k$ incurred by \texttt{BlockDot}$(A, B)$ satisfies: 
\begin{equation}
\label{eq:blockdot_bound}
    |E_k| \leq |X^{(A)} X^{(B)}| (k-1) u \sum_{i=1}^{k} 
    |P_i^{(A)} P_i^{(B)}| + O(u^2)
\end{equation}
The multiplications $P_i^{(A)} P_i^{(B)}$ can be performed exactly in fp32 for E4M3 and E5M2 elements, and the final scale multiplication 
$X^{(A)} X^{(B)}$ is exact since both are powers of two (assuming no overflow or underflow). Hence, no rounding error is introduced by these multiplications; the bound reflects only the accumulation error, scaled by the shared scale factors.

\paragraph{GeneralDotBounds}
\texttt{GeneralDot} takes two input vectors of length $n$ and splits them each into $k$ sized blocks. It then calls \texttt{BlockDot} on every corresponding pair of blocks (see Equation~\ref{eq:dotgeneral}). In these tests, input error introduced by the initial conversion to MX format is removed, so that the bound reflects only the accumulation error. Hence, the bound is based on the accumulation error. The accumulation precision used is also equivalent to the output precision.

Substituting into \cite[eq.~4.4]{higham2002} and including the per-block errors from Equation~\ref{eq:blockdot_bound}, the error 
$E$ incurred by \texttt{GeneralDot}$(A, B, MX)$ satisfies:

\begin{equation}
\label{eq:generaldot_bound}
\begin{split}
    |E| & \leq \ (\lceil n/k \rceil - 1) u \Bigg(
    \sum_{j=1}^{\lceil n/k \rceil} |A_jB_j| \\
    & + (k-1) u \sum_{j=1}^{\lceil n/k \rceil}
    |X^{(A_j)} X^{(B_j)}| \sum_{i=1}^{k} |P_i^{(A_j)} P_i^{(B_j)}| \Bigg) \\
    & + O(u^2)
\end{split}
\end{equation}
where $A_j$ and $B_j$ are the j-th blocks of
$A$ and $B$ respectively after MX conversion.

\begin{figure}[t]
    \centering
    \includegraphics[width=\columnwidth]{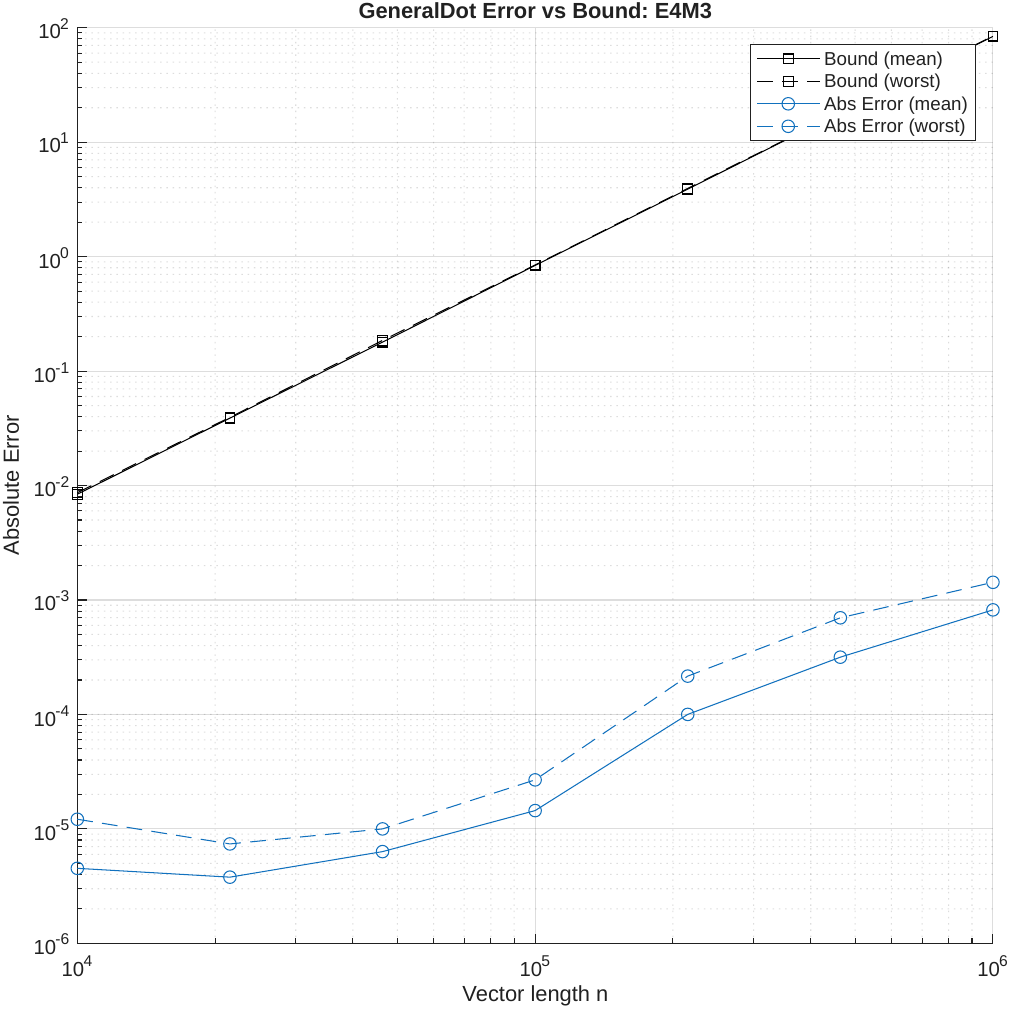}
    \caption{Absolute error and theoretical bound for 
    \texttt{GeneralDot} across vector lengths from $10^4$ to $10^6$, 
    for E4M3 element formats. Mean and worst-case values over 5 trials are shown.}
    \label{fig:generaldot_bound0}
\end{figure}

\begin{figure}[t]
    \centering
    \includegraphics[width=\columnwidth]{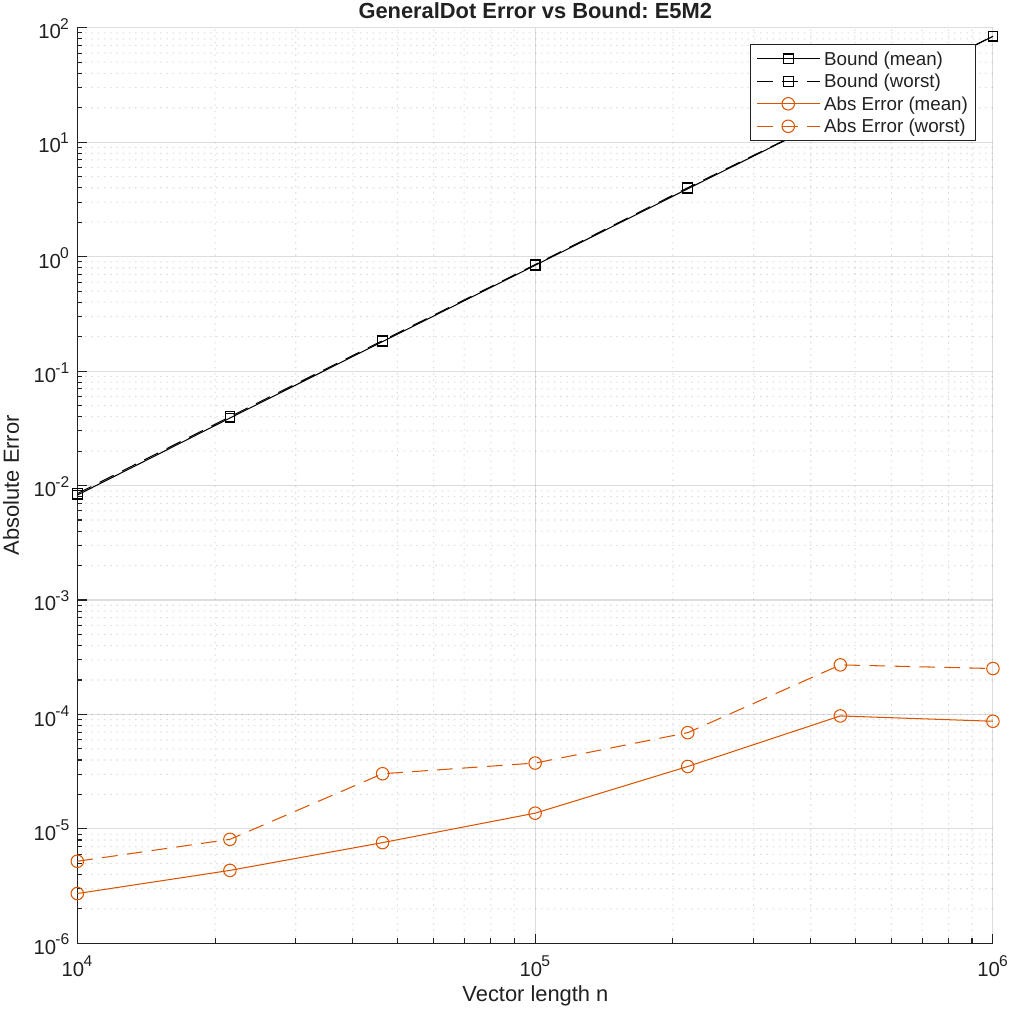}
    \caption{Absolute error and theoretical bound for 
    \texttt{GeneralDot} across vector lengths from $10^4$ to $10^6$, 
    for E5M2 element formats. Mean and worst-case values over 5 trials are shown.}
    \label{fig:generaldot_bound1}
\end{figure}

Figures~\ref{fig:generaldot_bound0}~and~\ref{fig:generaldot_bound1} show the results of plotting this bound against \texttt{GeneralDot} errors. All errors fall below the bound.

\section{Results}
The following example shows the configuration for simulating MXFP8 (E5M2) formats, with an output in fp16 and accumulation in fp32. 

\begin{verbatim}
MX.element_opts.format = 'E5M2';
MX.k = 32;
MX.accumulator_opts.format = 's'; 
MX.output_opts.format = 'h'; 
\end{verbatim}

The following example demonstrates \texttt{GeneralDot} usage, computing 
the dot product of two fp64 (binary64 IEEE 754 format) vectors of size $10^4$, using the MX format configuration described above. 

\begin{verbatim}
>> rng(500);
% Choose two random vectors
A = (rand(10^4,1) - 0.5) * 2;
B = (rand(10^4,1) - 0.5) * 2;
>> GeneralDot(A,B,MX)

ans =
    4.4375
\end{verbatim}

The following example demonstrates \texttt{BlockMatMul} usage with the 
same configuration.

\begin{verbatim}
>> rng(500);
>> C = (rand(4,4) - 0.5) * 2;
>> D = (rand(4,4) - 0.5) * 2;
>> BlockMatMul(C,D,MX)

ans =
    0.1094   -0.5820   -0.7812   -0.9897
   -0.0352   -0.3828   -0.5586    0.4756
    0.6416    0.4724    0.1628    0.2886
   -0.0371    0.0449   -0.0293   -0.3716
\end{verbatim}

\section{Conclusion}
We have presented MXsim v0.1, a MATLAB simulator for OCP MX block floating-point arithmetic, supporting the standard OCP operations with 
full customisation of element format, accumulator format, output format, block size, rounding and scale bits. In the future work
we plan to extend MXsim to support the IEEE P3109 standard
\cite{P3109_Public}, incorporate accurate models of NVIDIA Tensor Core accumulation \cite{khattak2025tensorcore}, and add fixed-point
accumulation support, for error-free accumulation.
The software \texttt{MXSim v0.1} is available on GitHub.\footnote{\url{https://github.com/north-numerical-computing/MXsim}}

\balance
\bibliographystyle{IEEEtran}
\bibliography{IEEEabrv,references}

\end{document}